\documentclass[5p]{elsarticle}     

\usepackage{graphicx}
\usepackage{caption}
\usepackage{subfigure}
\usepackage{subfloat}
\usepackage{stfloats}  
\usepackage{float}
\usepackage{booktabs}
\usepackage{multirow}
\usepackage{amsmath,amssymb,latexsym}
\usepackage{lineno,hyperref}
\modulolinenumbers[5]

\begin{document}
	
\begin{frontmatter}
\title{A deep learning based known-plaintext attack method for chaotic cryptosystem}
%
\author[1,2]{ Fusen Wang }
\author[1,2]{ Jun Sang\corref{cor1} }
\ead{ jsang@cqu.edu.cn }
\cortext[cor1]{Corresponding author}
\author[1,2]{ Qi Liu }
\author[1,2]{ Chunlin Huang}
\author[1,2]{ Jinghan Tan}
\address[1]{  Key Laboratory of Dependable Service Computing in Cyber Physical Society of Ministry of Education, Chongqing University, Chongqing 400044, China }
\address[2]{School of Big Data \& Software Engineering, Chongqing University, Chongqing 401331, China}

\begin{abstract}
In this paper, we propose a known-plaintext attack (KPA) method based on deep learning for traditional chaotic encryption scheme. 
We employ the convolutional neural network to learn the operation mechanism of chaotic cryptosystem, and accept the trained network as the final decryption system.
To evaluate the attack performance of different networks against different chaotic cryptosystem, we adopt two neural networks to perform known-plaintext attacks on two distinct chaotic encryption schemes.
The experimental results demonstrate the potential of deep learning-based method for known-plaintext attack against chaotic cryptosystem.
Different from the previous known-plaintext attack methods, which were usually limited to specific chaotic cryptosystems, a neural network can be applied to the cryptanalysis of various chaotic cryptosystems with deep learning-based approach, while several different networks can be designed for the cryptanalysis of chaotic cryptosystems. 
This paper provides a new idea for the cryptanalysis of chaotic image encryption algorithm.
\end{abstract}

\end{frontmatter}
\section{Introduction}
\label{sec:intro}

Many chaotic image encryption algorithms have been proposed during the past years, and their security requirements have achieved good improvement \cite{1,2,3,4,5,6}. 
While the chaotic cryptosystem is developing rapidly, its security has also attracted widespread attention. 
Many attack schemes against certain specific chaotic encryption algorithm have been proposed in succession, e.g., Dou et al. \cite{7} proposed an effective attack scheme against the one-dimensional combined chaotic color image encryption algorithm \cite{8} in the case of unknown parameters. 
Li et al. \cite{9} proposed a chosen-plaintext attack strategy against one-dimensional bit-level chaotic color image encryption algorithm \cite{10} and proved the effectiveness and feasibility of the method. 
However, the disadvantage of these methods is that they are limited to solving particular chaotic encryption schemes, and cannot attack other encryption algorithms.

In recent years, deep learning also has numerous applications in the domain of cryptography, especially in image encryption \cite{11,12}. 
Besides, Hai et al. \cite{13} first put forward one method to apply deep learning to the cryptanalysis of optical encryption scheme. 
They designed a deep neural network namely DecNet to train iterative a large number of ``plaintext-ciphertext'' pairs generated by the classical double random phase encoding system even the more secure triple random phase encoding system, and then regarded the trained model as the ``equivalent key'' of the cryptographic system, which was used to crack the subsequent ciphertext.

Based on the shortcomings of traditional chaotic system cryptanalysis and the stimulation of \cite{13} method, in this paper, we propose a known-plaintext attack method based on deep learning for chaotic cryptosystems.
Two chaotic encryption schemes are adopted to encrypt images to generate sufficient ``plaintext-ciphertext'' pairs, and two convolution neural networks are utilized as the decryption model to train these data. Then ciphertext images are performed to verify and visualize the decryption effect on the trained model.

The main contributions of our work are outlined as follows:
\begin{itemize}
  \item A known plaintext attack method based on deep learning for chaotic cryptosystems is proposed for the first time.
  \item By employing two different convolution neural networks to train the ``plaintext-ciphertext'' pairs, which are generated by two kinds of chaotic encryption schemes, excellent ciphertext reconstruction results can be obtained.
  \item We get the conclusion, different from the traditional known-plaintext attack methods for chaotic cryptosystems, one neural network can be employed to decrypt different chaotic cryptosystems, and one chaotic cryptosystem can take multiple neural networks as decrypters for comparing their ciphertext reconstruction effects.
\end{itemize}

\section{CNN for known-plaintext attack on chaotic cryptosystems}
The decryption work of ciphertext images is quite similar to some classic tasks in deep learning, such as image denoising \cite{14}, image defogging \cite{15} and image super-resolution reconstruction \cite{16}, in which their purposes are always to recover some fuzzy or even invisible images to the same resolution as the original image as much as possible. 
Therefore, we adopt two encoder-decoder neural networks based on end-to-end structure as experimental models, including the medical segmentation network Unet \cite{17} that is the earliest used for image denoising tasks, and the Multi-Stage Encoder-Decoder Network (MSEDNet) proposed by us. The specific introduction is as follows.

\subsection{Unet}
\begin{figure}[ht]              
	\centering	
	\includegraphics[scale=0.4]{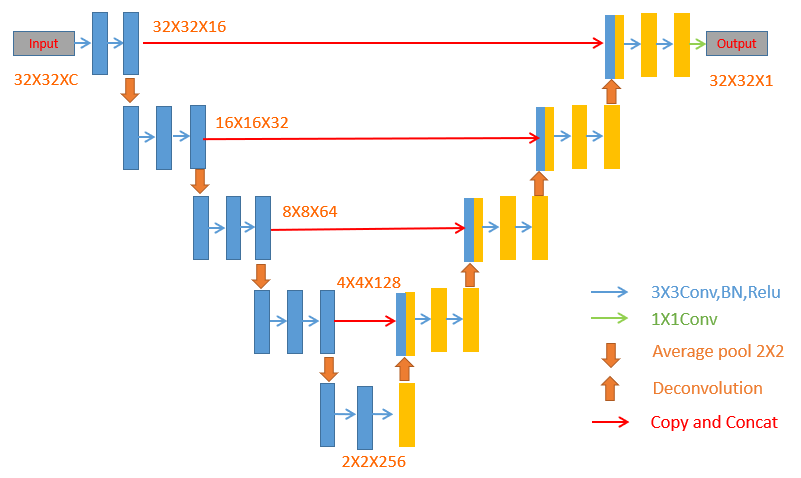}
	\caption{The architecture of modified Unet: H x W x C represents respectively the image' height, width, and channel number with a value of 1 or 3.}
	\label{img1:Unet}
\end{figure}

As a classic encoder-decoder network in the field of medical image processing, Unet \cite{17} is usually conducted for image segmentation, image compression, image denoising, etc. It can also be used to decrypt encrypted information and restore images. A straightforward modification is made to Unet' output channels since the datasets contain vast small-size images, and its network structure is shown in Fig. \ref{img1:Unet}. 
 
In addition to the last convolution layer with kernel size 1x1, each convolution consists of three consecutive operations: convolution (Conv) with 3x3 kernel size, batch normalization (BN), and rectified linear unit (ReLU). 
The down-sampling phase adopts the max-pooling of stride 2x2.  The up-sampling layers are concatenated with the output of the corresponding down-sampling layer and deconvolved with a 2x2 filter.

\subsection{MSEDNet}
In this paper, we propose an encoder-decoder network for decryption, named Multi-Stage Encoder Decoder Network. The first five layers of the left column of the network are encoding layers, and each layer is composed of convolution (Conv) with filter size 3x3, batch normalization (BN), rectified linear unit (ReLU), and average pooling layer with stride 2x2.
Through feature extraction, higher-precision pixel-by-pixel estimation can be achieved while maintaining the per-pixel position information on the feature map. 
 
\begin{figure}[ht]              
	\centering	
	\includegraphics[scale=0.4]{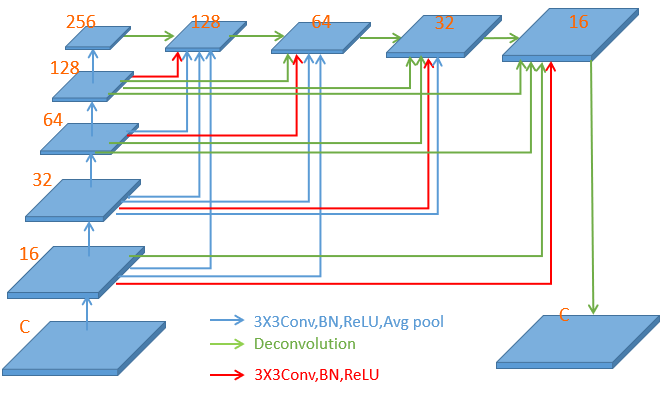}
	\caption{The architecture of MSEDNet: The numbers in the figure represent the number of channels in each feature map.}
	\label{img2:MSEDNet}
\end{figure}

 In the decoding stage, we integrate the feature map of multiple encoding stages to fully fuse the multi-scale context information and conduct pixel-by-pixel regression considering that simple interpolation operation cannot restore the information lost in the previous pooling layer, even if skip connections are employed. The network structure is presented in Fig. \ref{img2:MSEDNet}.

\vspace{-0.5cm}

\section{Chaotic encryption system}
To better verify that the known-plaintext attack method based on deep learning can be applied to the cryptanalysis of different chaotic encryption systems, we introduce two chaotic encryption schemes as the targets to be attacked: (1) the classic one dimensional chaotic map (for grayscale images); (2) the proposed hybrid chaotic map, i.e., employ one dimensional Logistic, Sine, and Chebyshev maps to encrypt the R, G and B channels of the image separately (for color images).

\subsection{One-dimensional chaotic maps}
This part briefly introduces three types of classic one-dimensional chaotic maps: Logistic map, Sine map, and Chebyshev map \cite{18}. In view of the space limitation and the similar nature of the experiment, we only employ Logistic map as the experimental scheme in one-dimensional chaotic map.

\subsubsection{Logistic map}
One-dimensional Logistic map has extremely complex dynamic behavior, and its mathematical expression formula is as follows:

\begin{equation}
	{X_{{\rm{n}} + 1}} = \mu  \times {X_n} \times (1 - {X_n}),
\end{equation}
where ${X_n}$ is the nth chaotic iterative sequence with vector length n. The initialization parameters ${X_0} \in \left( {0,1} \right)$, $\mu  \in \left[ {0,4} \right]$, and when $\mu  \in \left( {3.5699456,4} \right]$ is especially closer to 4, the iteration sequence will be in a pseudo-random distribution state. In experiment, we set $X_0=0.1$, $\mu = 3.601$.

\subsubsection{Sine map}
As one of classical one-dimensional chaotic maps, Sine map can generate chaotic sequences similar to Logical map. It can be defined by the following equation:

\begin{equation}
	{X_{n + 1}} = \sigma  \times \sin (\pi  \times {X_n}),
\end{equation}
where ${X_n}$ is the output chaotic iterative sequence, the initial parameter $\sigma  \in \left( {0,1} \right]$. We set $\sigma  = 0.95$, $X_0=0.154$.

\subsubsection{Chebyshev map}
Chebyshev map, which is similar to sine map, displays some specific chaotic effects. This function is formulated as below:
\begin{equation}
	{X_{n + 1}} = \cos (\theta  \times {\cos ^{ - 1}}{X_n}),
\end{equation}
when the parameter $\theta  > 1$, the output sequence has chaotic behavior. In our experiment, we set $\theta  = 5$, $X_0=0.165$.

\subsection{Hybrid chaotic maps}
This section will introduce the above-mentioned one-dimensional hybrid chaotic map.
In \cite{6}, one novel image encryption scheme was proposed to combine two different chaotic maps and derive a new encryption formula, which was used to encrypt the image and reduce the correlation between the three channels of image R, G and B.
Based on this idea, three different one-dimensional chaotic maps are adopted in this paper to perform chaotic processing on R, G and B channels of the color image.
Compared with using only one-dimensional Logistic map to generate a sequence of length W x H x C (W, H, C denotes width, height, channels of the image), the hybrid chaotic map is more secure and can further reflect the ability of convolution neural network to learn the encryption mechanism of chaotic systems.
In Section 5, we observe that the hybrid chaotic map is better than the one-dimensional Logistic map in terms of the chaotic effect through the evaluation metrics in Fig. \ref{img3:vis_mnist} and Fig. \ref{img5:vis_Cifar}.
The mathematical formula is as follows:

\vspace{-0.1cm}
\begin{equation}
	{Y_i} = Y(:,:,i){\rm{    }}i = 1,2,3,
\end{equation}
\vspace{-0.5cm}
\begin{equation}
	{Y_{1,n + 1}} = \mu  \times {Y_{1,n}} \times (1 - {Y_{1,n}}),
\end{equation}
\vspace{-0.5cm}
\begin{equation}
	{Y_{2,n + 1}} = \sigma  \times \sin (\pi  \times {Y_{2,n}}),
\end{equation}
\vspace{-0.5cm}
\begin{equation}
	{Y_{3,n + 1}} = \cos (\theta  \times {\cos ^{ - 1}}{Y_{3,n}}),
\end{equation}
\vspace{-0.1cm}
where Eq.(4) indicates that the color image $Y$ is divided into three single-channel images $Y_i$ along channel; in Eq.(5), Eq.(6), Eq.(7), $Y_{i,n+1}$ represents the chaotic processing of the R, G, and B channels of the color image $Y$ with Logistic map, Sine map, and Chebyshev map respectively.

\section{Implementation Details }
In this section, we give the experimental details of known-plaintext attacks on different chaotic cryptosystems by various convolution neural networks.
It is assumed that the attacker has obtained a large number of ``plaintext-ciphertext'' pairs without knowing the encryption algorithm since the neural network needs abundant data samples to learn the map relationship between the input end and the output end.
So, through two encryption schemes, we produce firstly sufficient ``plaintext-ciphertext'' pairs on datasets, which is then transmitted to two neural networks as training sets.
In the end, the ciphertext images of the testing set are input into the trained decryption model to verify the reconstruction effect.

\subsection{Datasets}
In this paper, we generate the ``plaintext-ciphertext'' pairs required for the experiment on MNIST \cite{19} and CIFAR-10 \cite{20}.
Since MNIST are all single-channel images, we employ an one-dimensional Logistic map to encrypt the datasets, while utilize hybrid chaotic map to encrypt the three-channel images of cifar-10.
These ``plaintext-ciphertext'' pairs are randomly divided, 90$\%$ of which are used as training sets of neural networks, while the rest are employed as testing sets to detect the decryption ability of model.

\subsection{Loss Function}
To minimize the information loss between the decrypted images and the plaintext images, the absolute value loss function (L1loss) is used to supervise the training of all the networks in this paper.
The equation is defined as follows:

\vspace{-0.2cm}
\begin{equation}
	Los{s_1} = \frac{1}{N}\sum\nolimits_{i = 1}^N {O({X_i};\theta ) - P({X_i})}, 
\end{equation}
where $N$ is the number of the training images batch, ${O({X_i};\theta )}$ is the decrypted image output by the network with parameters ${\theta}$, ${P({X_i})}$ is the supervision label (plaintext) of the ciphertext ${X_i}$.

\subsection{Training details}
All experiments are implemented in the environment of CPU-Python3.5-Pytorch1.0, employing Adam optimizer with L2 regularization weight decay rate of 1e-4.
The initial learning rate is 1e-5 and it drops by 10\% every 20 epochs when training with a batch size of 8. 
In addition, a dropout layer with ratio of 0.5 is added to the partial convolution layer to prevent the network from over-fitting. 

\section{Experiments}

\subsection{Evaluation metrics}
 In the experiment, Pearson correlation coefficient \cite{21} was taken as the evaluation metrics for two aspects: (1) to estimate the correlation between plaintext images and ciphertext images in different chaotic encryption systems in Section3; (2) to evaluate the correlation between the output (decryption image) and the label (plaintext image) of the known-plaintext attack method based on deep learning.
 It can be defined by the following equation:
 
\begin{equation}
	E(X) = \frac{1}{{W \times H}}\sum\nolimits_{i = 1}^W {\sum\nolimits_{j = 1}^H {X(i,j)} }, 
\end{equation}
\vspace{-0.5cm}
\begin{equation}
	\sigma (X) = \sqrt {\sum\nolimits_{i = 1}^W {\sum\nolimits_{j = 1}^H {{{[X(i,j) - E(X(i,j))]}^2}} } }, 
\end{equation}
\vspace{-0.5cm}
\begin{equation}
	Corr = \frac{{\sum\nolimits_{i = 1}^W {\sum\nolimits_{j = 1}^H {(O - E(O))(P - E(P))} } }}{{\sigma (O)\sigma (P)}},
\end{equation}
where $W$, $H$ represent the width and height of the images respectively, $O(i,j)$ is the output of the network (decrypted image), $P(i,j)$ is the label of the network (plaintext image). This Eq.(11) is applicable to two encryption schemes and is regarded as training/testing accuracy.

\subsection{Experiment results of Logistic map on MNIST}
Table. \ref{table1} shows the experimental results of two convolution neural networks to crack the Logistic map on the MNIST dataset, including training Loss, training average accuracy, testing average accuracy, epoch and the running time of each epoch.

\begin{figure}[htp]
	\vspace{-1.0cm}             
	\centering	
	\includegraphics[scale=0.4]{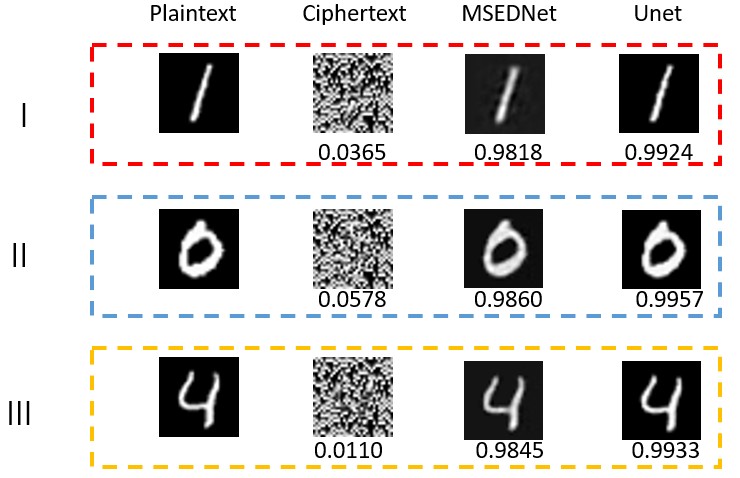}
	\caption{Visualization of three random images on MNIST testing set, including Plaintext, Ciphertext, Decrypted image of MSEDNet, Unet, Correlation coefficient with plaintext.}
	\label{img3:vis_mnist}
\end{figure}

Fig. \ref{img3:vis_mnist} gives the visualization results of one-dimensional Logistic map on the MNIST dataset, where the first column shows the plaintext image; the second column displays the corresponding ciphertext image; the third-fourth columns list the decrypted images of MSEDNet, and Unet severally.
At the same time, the correlation coefficients of the ciphertext image and the plaintext image, as well as the decrypted images and the plaintext image are given separately. From the results we can observe that the testing ciphertext images can be well reconstructed by the neural network, and the correlation with the plaintext gradually tends to 1. However, the reconstruction effect of diverse networks is different, among which the decryption effect is Unet > MSEDNet.

\begin{figure}[htp]
	\vspace{-1.1cm}
	\begin{minipage}[t]{1.0\linewidth}
		\centering
		\includegraphics[height=5.0cm,width=7.5cm]{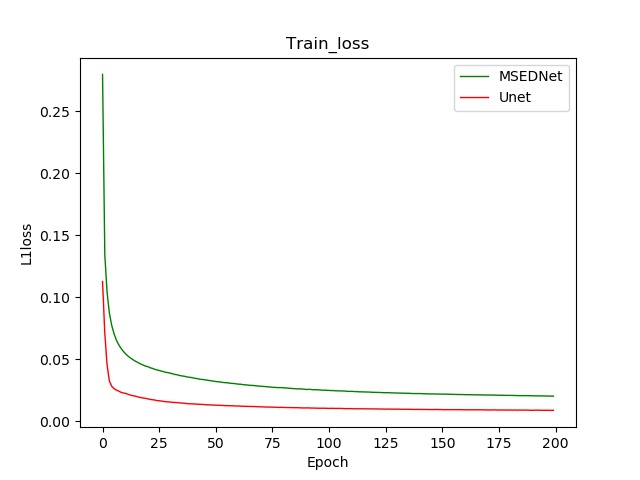}
		\centerline{(a)}\medskip
	\end{minipage}

	\begin{minipage}[t]{1.0\linewidth}
		\centering
		\includegraphics[height=5.0cm,width=7.5cm]{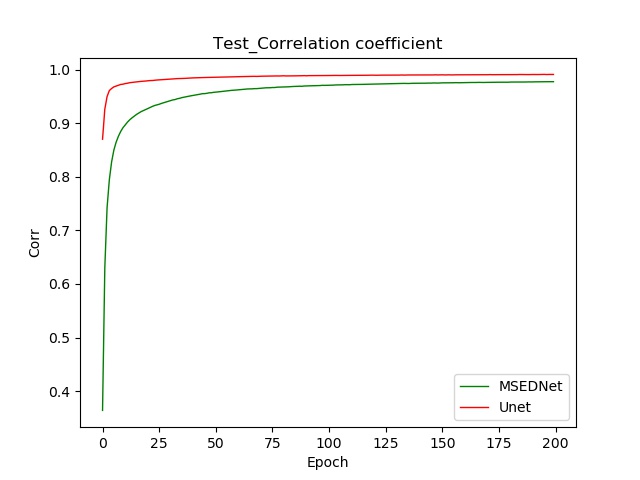}
		\centerline{(b)}\medskip
	\end{minipage}		
	\caption{ The change curve of MNIST training process: (a) Training L1Loss, (b) Testing Correlation Coefficient. The green and red line represent MSEDNet and Unet architecture separately.}	
	\label{img4}                    
\end{figure}


The curve changes of training Loss and testing evaluation metrics of two networks on the MNIST dataset are plotted in Fig. \ref{img4} (a) and Fig. \ref{img4} (b). We reach the conclusion from the figure that the selections of neural network for the same encryption scheme greatly affect the training time and accuracy. The training efficiency and accuracy of Unet [17] are better than the MSEDNet.

\subsection{Experiment results of Hybrid chaotic map on Cifar}
Table. \ref{table2} presents the experimental results of different convolution neural networks cracking the hybrid chaotic map (Cifar-10), similar to Table. \ref{table1}.

\begin{figure}[htp]              
	\centering	
	\includegraphics[scale=0.4]{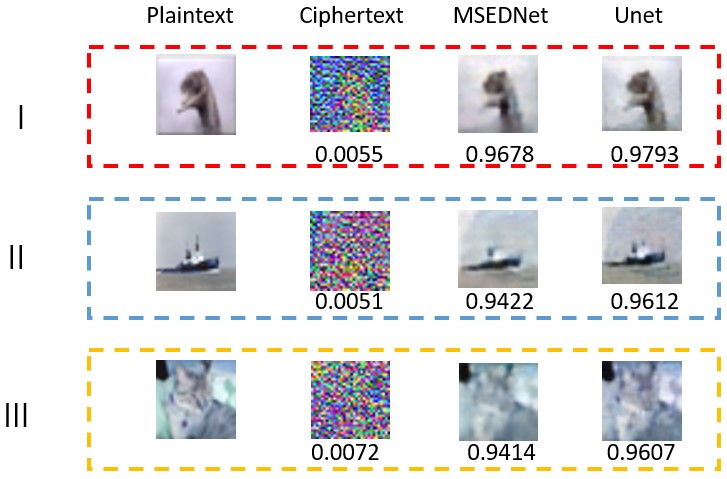}
	\caption{Visualization of three random images on Cifar-10 testing set, including Plaintext, Ciphertext, Decrypted image of MSEDNet, Unet, Correlation coefficient with plaintext.}
	\label{img5:vis_Cifar}
	\vspace{-0.3cm}
\end{figure}

\begin{table*}[htp]
	\centering
	\caption{The experimental result of deep learning method for attacking Logistic map.}	
	\begin{tabular}{cccccc}
		\toprule  
		\multicolumn{1}{l}{Ëncryption Scheme} & \multicolumn{1}{l}{Network} & \multicolumn{1}{l}{Training accuracy} &  \multicolumn{1}{l}{Testing accuracy} &  \multicolumn{1}{l}{Epoch} & \multicolumn{1}{l}{Time/Epoch} \\
		\midrule  
		
		\multirow{2}{*}{Logistic Map}       & MSEDNet                     & 98.32\%                               & 97.80\%                             & \multirow{2}{*}{200}              & 80s                            \\
		& Unet                        & 99.61\%                               & 99.15\%                              &                           & 85s                    \\
		\bottomrule 
	\end{tabular}
	\label{table1}
\end{table*}

\begin{table*}[htp]
	\centering
	\caption{The experimental result of deep learning method for attacking Hybrid Chaotic map.}	
	\begin{tabular}{cccccc}
		\toprule  
		\multicolumn{1}{l}{Ëncryption Scheme} & \multicolumn{1}{l}{Network} & \multicolumn{1}{l}{Training accuracy} & \multicolumn{1}{l}{Testing accuracy} & \multicolumn{1}{l}{Epoch} & \multicolumn{1}{l}{Time/Epoch} \\
		\midrule  
		
		\multirow{2}{*}{Hybrid Chaotic Map}    	& MSEDNet                     & 95.86\%                               & 93.62\%                                 & \multirow{2}{*}{800}                           & 82s                            \\
		& Unet                        & 97.82\%                               & 95.70\%                              &                           & 89s                    \\
		\bottomrule 
	\end{tabular}
	\label{table2}
\end{table*}

\begin{figure}[htp]
	\vspace{-0.1cm}
	\begin{minipage}[t]{1.0\linewidth}
		\centering
		\includegraphics[height=5.0cm,width=7.5cm]{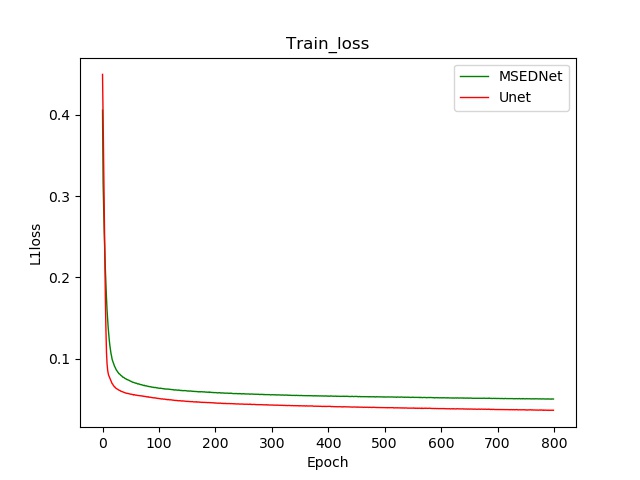}
		\centerline{(a)}\medskip
	\end{minipage}
	\vspace{-0.5cm}
	\begin{minipage}[t]{1.0\linewidth}
		\centering
		\includegraphics[height=5.0cm,width=7.5cm]{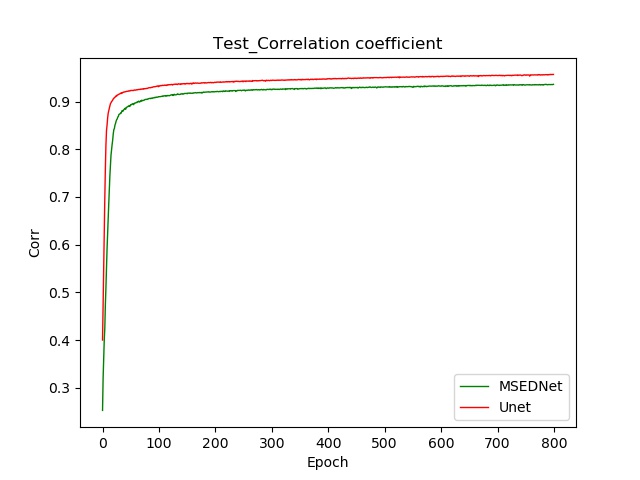}
		\centerline{(b)}\medskip
	\end{minipage}		
	\caption{ The change curve of MNIST training process: (a) Training L1Loss, (b) Testing Correlation Coefficient. The green and red line represent MSEDNet and Unet architecture separately.}	
	\label{img6}                    
\end{figure}

Fig. \ref{img5:vis_Cifar} shows the visualization results of the hybrid chaotic map on the Cifar-10 dataset, which is greatly similar to MNIST, i.e., after more epochs training, the ciphertext image in testing set can be reconstructed pretty well, but the two reconstruction effects have slight differences, among which the decryption effect is Unet > MSEDNet.
Fig. \ref{img6} displays the curve changes of training Loss and testing evaluation metrics of two networks on the Cifar-10 dataset. Similar to Fig. \ref{img4}, it can be observed that the training efficiency and accuracy of Unet are significantly higher than MSEDNet.

\subsection{Experimental analysis}
This paper aims to investigate the cryptanalysis of performing known-plaintext attacks with multiple convolution neural networks against different chaotic encryption systems.
The following conclusions can be drawn from the above experimental data in the figure and table: (1) For the same encryption scheme, different neural networks can be used as the decryption model.
However, the reason for different ciphertext reconstruction effects may be related to the number of participants and learning ability of the neural network, which will be one of our next research works.
(2) The same convolution neural network can be used to attack different encryption systems, but the ciphertext reconstruction effect is also diverse, which may be related to the complexity of the encryption mechanism.
This will be our other future research works.

Compared with traditional known-plaintext attack methods specific to a certain chaotic cryptosystem, our scheme deep learning-based is more flexible and extensible.
Moreover, this method can be transferred to other chaotic cryptosystems, even to the field of non-chaotic encryption.

\section{Conclusion}
 In this paper, a known-plaintext attack scheme of chaotic system based on deep learning is proposed, which uses convolution neural networks to train a large number of ``plaintext-ciphertext'' pairs to learn the conversion process between plaintext and ciphertext.
 The trained model is regarded as the ``decryption formula''.
 By training different encryption systems with various convolutional neural networks, we obtain excellent decryption results.
 Overall, the diversity of ciphertext reconstruction effects may be related to the complexity of the encryption system, the number of parameters and learning capabilities of the neural network, the training time.
 This will be worthy of our further study.
 In practice, the advantages of this method are as follows: (1) It can completely avoid the recording and transmission of complex ciphertexts; (2) The chaotic cryptanalysis method based on deep learning can be introduced to other chaotic cryptographic systems even to the field of non-chaotic cryptography; (3) In addition, it also proposes a new research direction in the field of multimedia security, i.e., how to prevent cryptography attack means based on deep learning.
\section{References}
\bibliographystyle{elsarticle-num}
\bibliography{mybibfile}
\end{document}